\documentclass[a4paper]{aipproc}

\newcommand{\cd}{\makebox[0.08cm]{$\cdot$}}

\layoutstyle{6x9}
  
\begin{document}

\title[Stability of bound states in the]
{Stability of bound states in the light-front Yukawa model}

\classification{43.35.Ei, 78.60.Mq}
\keywords{Document processing, Class file writing, \LaTeXe{}}

\author{V.A. Karmanov}{
  address={Lebedev Physical Institute, Leninsky Pr. 53, 119991 Moscow, Russia },
  email={karmanov@sci.lebedev.ru},
}

\iftrue
\author{J. Carbonell}{
  address={Institut des Sciences Nucl\'{e}aires,
        53, Av. des Martyrs, 38026 Grenoble, France},
  email={carbonel@isn.in2p3.fr},
}

\author{M. Mangin-Brinet}{
  address={ Institut des Sciences Nucl\'{e}aires,
        53, Av. des Martyrs, 38026 Grenoble, France},
  email={mariane@isn.in2p3.fr},
}
\fi

\copyrightyear  {2001}

\begin{abstract}
We show that in the system of two fermions interacting by scalar  
exchange, the solutions for J$^{\pi}$=$0^+$ bound states are stable 
without any cutoff regularization, for values of the coupling constant 
$\alpha$ below a critical value $\alpha_c$.  This latter  
is calculated from an eigenvalue equation.
\end{abstract}

\date{\today}

\maketitle

\section{Introduction}
The Yukawa model, describing a system of two fermions interacting by scalar 
exchange  (${\cal L}^{int}=g\bar{\psi}\psi\phi^{(s)}$), is instructive 
for studying the relativistic bound states and for developing the 
renormalization methods. It is also a main ingredient in building the $NN$ 
interaction, which contains an important contribution of 
scalar meson exchanges.
The bound state problem and the renormalization in the Yukawa model 
were studied \cite{glazek} in the framework of standard light-front 
dynamics \cite{BPP_PR_98}.
The relativistic two-nucleon wave functions have been also 
calculated perturbatively \cite{ckj10} in the explicitly covariant 
version of light-front dynamics \cite{karm76}, where the state 
vector is defined on the invariant plane
$\omega\cd x=0$ with $\omega^2=0$ (see for review \cite{cdkm}).
In this work, the Bonn $NN$ model was used with the
corresponding form factors \cite{Bonn} and the problem of
cut-off dependence was not analyzed.

In reference \cite{Mariane}, we investigated the 
stability of the bound states relative to the high momentum 
contributions of the kernel, when the cutoff tends to infinity.  Below we present 
the results of our study and compare them with those 
obtained in \cite{glazek}.

\section{The cutoff dependence of the binding energy}
We consider the two fermion wave function with total angular 
momentum $J=0^+$.  In the fermion spin indices, it is a $2\times 2$ 
matrix determined, due to the parity conservation, by two independent 
elements -- the spin components $f_{i=1,2}$.  Since the wave 
function is defined on the light-front plane $\omega\cd x=0$, 
components $f_{i}(k,\theta)$ depend not only on the relative momentum 
$k$, but also on the angle $\theta$ between $\vec{k}$ and 
$\hat{n}=\vec{\omega}/|\vec{\omega}|$. The system of equations for 
$f_{i}$ contains a $2\times 2$ matrix kernel $K_{ij}$, 
calculated by using the explicitly covariant
light-front graph techniques \cite{karm76,cdkm}. These equations and 
the analytical expressions of kernels are given in \cite{Mariane}.
Though their form differs from the ones used in \cite{glazek}, 
we have shown that they are strictly equivalent. 

Let us consider the equations on the finite interval $0\leq k\leq k_{max}$.  
The dependence of their solutions on the cutoff $k_{max}$ in the limit 
$k_{max}\to \infty$ is determined by the kernels asymptotics.  The 
kernel $K_{22}$ is repulsive and cannot generate  a collapse, 
whereas $K_{11}$ is attractive.  Therefore, to investigate 
the stability, we can consider the one channel problem for the 
component $f_1$, which satisfies the equation:
\begin{equation}\label{as2_0}  
\left[M^2-4(k^2+m^2)\right] f_1(k,\theta) =\frac{m^2}{2\pi^3} \int
K_{11}(k,\theta;k',\theta')f_1(k',\theta')\frac{d^3k'}{\varepsilon_{k'}},
\end{equation}
where $\varepsilon_{k'}=\sqrt{m^2+k'^2}$.

Our analysis uses the fact 
that at $k\to\infty$ the integral in the 
r.h.s. of (\ref{as2_0}) is dominated by the region $k'\propto k$, 
i.e. $k'\to\infty$ with a fixed ratio $k'/k=\gamma$ \cite{smirnov}. 
One can therefore replace in (\ref{as2_0}) both wave function and 
kernel by their asymptotics, which have the form \cite{Mariane} :
\begin{equation}\label{eqn16}
f_1(k,z) = {h_1(z)\over k^{2+\beta}},
\qquad
K_{11}=-\frac{\pi\alpha}{m^2}\left\{
\begin{array}{ll}
\sqrt{\gamma}A_{11}(z,z',\gamma),
& \mbox{if $\gamma<1$}
\\
A_{11}(z,z',1/\gamma)/\sqrt{\gamma},
& \mbox{if $\gamma>1$}
\end{array}\right.
\end{equation}
with   $0< \beta< 1$ and
\begin{equation}\label{eq14a}
A_{11}(z,z',\gamma)=
\int_0^{2\pi}
\frac{d\phi}{2\pi\sqrt{\gamma}}\,
\frac{2\gamma(1-z z')-
(1+\gamma^2)\sqrt{1-z^2}\sqrt{1-z'^2}\cos\phi}
{(1+\gamma^2)(1+|z-z'|-z z') -2\gamma\sqrt{1-z^2}\sqrt{1-z'^2}\cos\phi},
\end{equation}
where $\alpha=g^2/4\pi$ and $z=\cos\theta$. By setting
$A_{11}(z,z',\gamma)\equiv 1$ and $\alpha=2\pi m\alpha'$ in (\ref{eqn16}), 
the asymptotics of $K_{11}$ becomes identical to asymptotics of 
the momentum space kernel corresponding to the non-relativistic 
potential $-\alpha'/r^2$.  As it is well known \cite{ll},  
there exists for this potential a critical coupling constant 
$\alpha_c'=1/(4m)$, that 
corresponds to $\alpha_c=\pi/2$.  The inspection of (\ref{eq14a}) shows 
that in the Yukawa model the function $A_{11}$ is smaller than for the 
$-\alpha'/r^2$  potential:  $0\leq A_{11}(z,z',\gamma) \leq 1$.  
Therefore in this model, one can expect a larger critical coupling 
constant i.e. $\alpha_c>\pi/2$, what is confirmed by numerical 
calculations.

Substituting (\ref{eqn16}) into equation (\ref{as2_0}), we obtain 
for $g_1(z)$ \cite{smirnov}:  
\begin{equation}\label{gHg}
\int_{-1}^{+1} dz' H_{\beta}(z,z')\;h_1(z')=\lambda h_1(z)
\end{equation}
with $\lambda=1/\alpha$ and 
\begin{equation}\label{H_beta}
H_{\beta}(z,z')=\int_0^1{d\gamma\over 2\pi\sqrt\gamma}\; 
A_{11}(z,z',\gamma)\;\cosh{(\beta\log\gamma)}.
\end{equation} 
Equation (\ref{gHg}) is an eigenvalue equation for $\lambda$, parametrized by
$\beta$. It provides the relation between the coupling constant $\alpha$ and  
$\beta$, determining the power law (\ref{eqn16}) of the wave 
function asymptotics. The r.h.s. of equation (\ref{as2_0}) becomes 
divergent for $\beta\leq 0$.  Hence, the equation $\beta(\alpha_c)=0$ 
determines the critical coupling constant $\alpha_c$.  

\section{Results}\label{num}
\begin{figure}[h]
\resizebox{16pc}{!}{\includegraphics[height=.5\textheight]{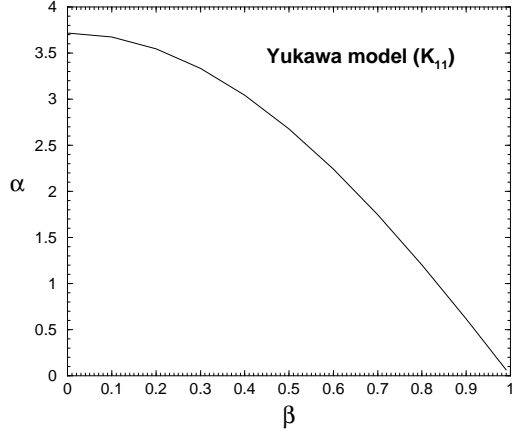}}
\caption{Function $\alpha(\beta)$ determined by eq. (\protect{\ref{gHg}}).
The critical coupling constant is: $\alpha_c=\alpha(\beta=0)=3.72$. The 
values discussed in the text: $\alpha=1.096,\beta=0.819$ and 
$\alpha=2.480,\beta=0.548$ are on the curve.} \label{alpha_beta}
\end{figure}

In the numerical calculations, the constituent masses were taken equal 
to $m$=1 and the mass of the exchanged scalar is $\mu$=0.25.  The 
numerical solution $\alpha(\beta)$, found from equation (\ref{gHg}) 
with the function $A_{11}(\gamma,z,z')$ given by (\ref{eq14a}), is 
plotted in figure \ref{alpha_beta}.  The critical coupling constant is 
obtained for $\beta=0$ for which the eigenvalue is $\lambda_c=0.269$.  
It corresponds to $\alpha_c=1/\lambda_c=3.72$, as shown in figure 
\ref{alpha_beta} at $\beta=0$.

\begin{figure}[h]
\resizebox{16pc}{!}{\includegraphics[height=.45\textheight]{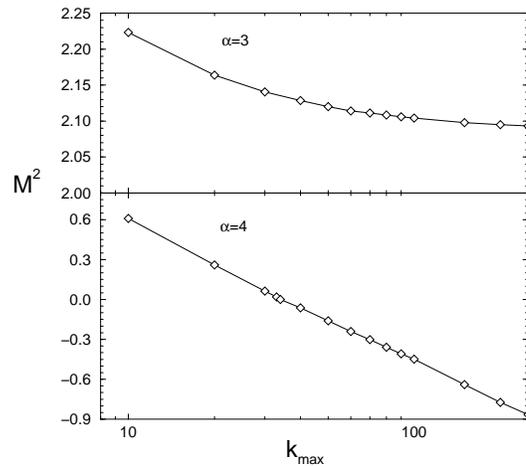}}
\caption{Cutoff dependence of the binding energy in the $J=0^+$ 
state, in the one-channel problem ($f_1$), for two fixed 
values of the coupling constant below and above the critical 
value.}\label{B_kmax} 
\end{figure}

In figure \ref{B_kmax}, we have plotted the mass square $M^2$ of the two 
fermion system, found from (\ref{as2_0}), as a function of the cutoff 
$k_{max}$ for two fixed values of the coupling constant below and above  
the critical value $\alpha_c=3.72$.  
One can see two dramatically different behaviors depending on the value 
of the coupling constant $\alpha$. For $\alpha=3$, i.e. 
$\alpha<\alpha_c$, the result is convergent. On the contrary, for 
$\alpha=4$, i.e. $\alpha>\alpha_{c}$, the result is clearly divergent: 
$M^2$ decreases logarithmically as a function of $k_{max}$ and becomes 
even negative.  Though the negative values of $M^2$ are physically 
meaningless, they are formally  allowed by equations.  The first degree 
of $M$ does not enter neither in the equation nor in the kernel, and 
$M^2$ crosses zero without any singularity.  

\begin{figure}[h]
\resizebox{16pc}{!}{\includegraphics[height=.5\textheight]{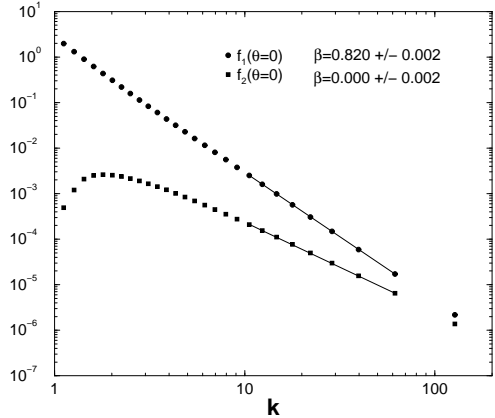}}
\caption{Asymptotic behavior of the $J=0^+$ wave function components 
$f_{1,2}$ for $B$=0.05, $\alpha$=1.096, $\mu$=0.25. The slope coefficients 
are $\beta_1=0.820$ and  $\beta_2\approx0$.}\label{wf_as_50}
\end{figure}

We have examined the asymptotical behavior of the wave function 
$f_1(k,z)$ and found that it very accurately follows the  power law 
(\ref{eqn16}) with the power $\beta(\alpha)$ given in figure 
\ref{alpha_beta}.  For instance for a binding energy $B=2m-M=0.05$ 
($\alpha=1.096$) a direct measurement in the numerical solution  
plotted in figure \ref{wf_as_50} gives $\beta=0.820\pm0.002$ whereas 
the solution of equation (\ref{gHg}) for the corresponding $\alpha$ 
gives $\beta=0.819$.  
The same kind of agreement was found for $B=0.5$ ($\alpha=2.480$, 
$\beta=0.548\pm 0.002$). 

We conclude that the $J=0$ -- or $(1+,2-)$ state in the classification 
\cite{glazek} -- is stable (i.e.  convergent relative to the cutoff 
$k_{max}\to \infty$) for coupling constant $\alpha$ below the critical 
value $\alpha_c=3.72$.  In this point, our conclusion differs from the 
one settled in \cite{glazek}, where it was stated that the integrals in 
the r.h.s. of the equations diverge logarithmically with cutoff.  
Above the critical value the integrals indeed diverge and the system 
collapses.

In the $J=1$ state the system is found to be always unstable, as 
pointed out in  \cite{glazek}. 

Thus, by an analytical method, confirmed by numerical calculations,
we have shown that the Yukawa model is 
not cutoff dependent for coupling constant below a critical value. 
The results obtained should be taken into account 
for instance in the renormalization procedures.
$ $ 
\vspace{-0.6cm}


\end{document}